\documentclass[prd,twocolumn,nofootinbib]{revtex4-2}
\usepackage{bm,url,natbib,textcase,color,hyperref,amsmath,amssymb,graphicx}
\usepackage[utf8]{inputenc}

 \newcommand{\on}{\operatorname}

\newcommand{\beq}{\begin{equation}}
\newcommand{\eeq}{\end{equation}}

\newcommand{\arXiv}[1]{\href{http://www.arXiv.org/abs/#1}{arXiv:#1}}
\renewcommand{\doi}[2]{\href{http://dx.doi.org/#2}{#1}}

\newcommand{\sst}{\scriptscriptstyle}

\newcommand{\dd}{{\rm d}}

\begin{document}

\title{Democracy from topology}

\author{Oleg Evnin}
\email{oleg.evnin@gmail.com}
\affiliation{High\ Energy\ Physics\ Research\ Unit, Faculty of Science, Chulalongkorn University,
Bangkok 10330, Thailand}
\vskip 3mm
\affiliation{Theoretische Natuurkunde, Vrije Universiteit Brussel (VUB) and\\ International
  Solvay Institutes, Brussels 1050, Belgium}
  
\author{Euihun Joung}
\email{euihun.joung@khu.ac.kr}
\affiliation{Department of Physics, College of Science, Kyung Hee University, Seoul 02447, Republic of Korea}
\author{Karapet Mkrtchyan} 
\email{k.mkrtchyan@imperial.ac.uk}
\affiliation{Theoretical Physics Group, Blackett Laboratory, Imperial College London, SW7 2AZ, UK}

\begin{abstract}

Chiral form fields in $d$ dimensions can be effectively described as edge modes of topological Chern-Simons theories in $d+1$ dimensions. At the same time, manifestly Lorentz-invariant Lagrangian description of such fields directly in terms of a $d$-dimensional field theory is challenging and requires introducing nontrivial auxiliary gauge fields eliminated on-shell with extra gauge symmetries. A recent work by Arvanitakis et al.\ demonstrates (emphasizing the case of 2d chiral bosons) that the two approaches are related, and a peculiar reduction on the $(d+1)$-dimensional topological Lagrangian automatically leads to $d$-dimensional Lagrangians with appropriate sets of auxiliary fields. We develop this setup in three distinct directions. First, we demonstrate how arbitrary Abelian self-interactions for chiral forms can be included using nonlinear boundary terms in the Chern-Simons theory. Second, by generalizing the Chern-Simons theory to the BF theory, we obtain an analogous democratic description of non-chiral 
form fields, where electric and magnetic potentials appear as explicit dynamical variables. Third, we discuss the effects of introducing
topological interactions in the higher-dimensional bulk, which produce extra interaction terms in the boundary theory. When applied to a topological 4-form field in 12 dimensions, this construction results in a democratic description of the 3-form gauge field of the 11-dimensional supergravity.

\end{abstract}

\maketitle

\section{Introduction}

It has been known for a long time \cite{Moore:1989yh,Witten:1996hc,Maldacena:2001ss,Losev:1995cr,Costello:2019tri} that the topological Chern-Simons theory and its BF generalizations can describe 
(chiral) $p$-form degrees of freedom on the boundary.
However, the generality and systematics of
this approach is not fully 
understood yet.

While the description of chiral fields as edge modes of topological theory is graceful and simple, the fact that one inevitably starts in a fictitious spacetime of one dimension higher may be seen as a drawback. Attempts to describe chiral fields as Lagrangian theories without introducing extra dimensions, on the other hand,
have met difficulties of their own. Early ventures in this direction sacrificed manifest Lorentz invariance \cite{Floreanini:1987as, Henneaux:1988gg,Perry:1996mk}.
The elegant Pasti-Sorokin-Tonin (PST) approach \cite{Pasti:1996vs,Pasti:1997gx,Buratti:2019guq} offers an economical Lorentz-invariant formulation, but suffers from
non-polynomial dependence of the action on an auxiliary scalar field, and furthermore encounters difficulties when including self-interactions \cite{Buratti:2019guq}.
(We mention additionally the approach of \cite{Sen:2019qit}, where chiral fields are necessarily accompanied by decoupled but propagating additional degrees of freedom. See also \cite{Lambert:2023qgs,Hull:2023dgp}.)

Recently \cite{Avetisyan:2022zza}, Lorentz-covariant Lagrangians for arbitrary self-interacting chiral $p$-forms were found.
The description includes a doubled set of gauge fields and an auxiliary scalar, which are gauged on-shell to a single propagating self-interacting chiral $p$-form.
A comparison of this formalism with other approaches in the literature can be found in \cite{Evnin:2022kqn}.

The topological field theory approaches to chiral forms have been pursued historically rather independently of the line of research
that builds Lagrangian descriptions of chiral forms using auxiliary fields without introducing extra spacetime dimensions.
A bridge connecting the two approaches was set up in a recent work by Arvanitakis et al.\ \cite{Arvanitakis:2022bnr} who found a reduction procedure\footnote{The reduction procedure of \cite{Arvanitakis:2022bnr} assumes a topologically trivial bulk with a single boundary. The nontrivial features of the bulk theory on manifolds of more complicated topology (see, e.g., \cite{Porrati:2021sdc}) thus do not enter the game in this setting. 
We thank Massimo Porrati for emphasizing the importance of this point.} 
that allows deriving the boundary theory from the Chern-Simons theory in the bulk. The procedure naturally leads to a boundary theory in the form of \cite{Avetisyan:2022zza} (which, for the case of free forms, can be related to PST formulation \cite{Mkrtchyan:2019opf} by integrating out auxiliary gauge fields).

Our present purpose is to extend and generalize the formulation of \cite{Arvanitakis:2022bnr} in a few different directions. First, arbitrary Abelian self-interactions
can be introduced to the setup of \cite{Arvanitakis:2022bnr} by adding nonlinear boundary terms to the Chern-Simons action. One thus recovers the full scope
of self-interacting theories in \cite{Avetisyan:2022zza}. Second, the problem of Lagrangian description of chiral forms is often discussed side-by-side with the problem of `democratic' description of ordinary (non-chiral) forms, where the dual electric and magnetic potentials appear as explicit dynamical variables.
As we shall see, such democratic theories emerge from boundary reductions of the topological BF theory, a cousin of the Chern-Simons theory evoked in
\cite{Arvanitakis:2022bnr}. Finally, in the BF setup, it is possible to introduce topological interactions in the bulk. This, correspondingly, affects the boundary theory inducing self-interactions that essentially involve the gauge potential (as opposed to being expressible through the field strength alone). In this way, in particular, one obtains a democratic description of the self-interacting 3-form appearing in the 11-dimensional supergravity.

\section{Chiral fields}

Here, we give a short derivation similar to that undertaken in \cite{Arvanitakis:2022bnr} for free chiral forms, adding Abelian interactions.

The starting point is the Chern-Simons theory given by the action
\begin{align}
    S=\int_M H\wedge {\rm d}H\,
\end{align}
(for our purposes the overall factor aka Chern-Simons level does not have to be explicit) where $M$ is a $d+1=2p+3$ ($p$ is even) dimensional manifold with a  boundary $\partial M$ and $H$ is a $(p+1)$-form field. 

The variation of this Lagrangian contains a boundary 
term $\int_{\partial M} \delta H\wedge H$, which would be 
incompatible with the least action principle. To remedy for this inconsistency,
we add a boundary term $-\frac12 H\wedge \star H$ to the action to obtain
\begin{equation}
    S_{\rm\sst free}
    =\int_M H\wedge \dd H-
    \frac12 \int_{\partial M}H\wedge \star H\,.
\end{equation}
The variation is then
\begin{equation}
        \delta S_{\rm\sst free}
    =2\int_M \delta H\wedge \dd H
    -\frac12 \int_{\partial M}\delta H^+\wedge H^-\,.\label{del S free}
\end{equation} 
Here and in what follows, we use the shorthand notation 
\beq
H^\pm=H\pm \star H,
\eeq 
and the pullback of $H$ onto the boundary is denoted by the same symbol $H$.
Note that $\star$ shall denote throughout the  Hodge dual associated with
an arbitrary metric on the boundary with
Lorentzian signature (the bulk Hodge dual will not appear in the formalism we consider, hence no danger of confusion).

We may impose
the Dirichlet boundary condition,
$\delta H^+=0$
or the Neumann one $H^-=0$:
$H^+$ and $H^-$ play the roles of `position' and `momentum,' respectively.
The Neumann condition 
can be also viewed as
the dynamical equation
with respect to the boundary variation.
We shall take the latter 
point of view as it
is more convenient for
introducing interactions.

As discussed in \cite{Avetisyan:2022zza,Evnin:2022kqn}, general equations describing self-interactions of a chiral field are given as
\begin{align}
    H^-=f(H^+)\,,\qquad {\rm d} H=0\,,
    \label{int bd eq}
\end{align}
where $f:\Lambda^+\to \Lambda^-$ is an
antiselfdual form valued function of a selfdual variable (here $\Lambda^+$ and $\Lambda^-$ represent the space of selfdual and antiselfdual forms respectively).

In order to reproduce these equations, one can 
introduce a boundary term to the Chern-Simons theory, given by an arbitrary function of $H^+$ as
\begin{align}
    S=\int_M H\wedge {\rm d}H - \int_{\partial M} \frac12\, H \wedge \star H+g(H^+)\,.\label{CSbInt}
\end{align}
The function $g(H^+)$ is a 
top form function of the 
selfdual argument $H^+$. 
The addition
of $g(H^+)$
is analogous
to the addition of 
an arbitrary potential term
to a free Hamiltonian.
The bulk equations of motion stemming from the action \eqref{CSbInt} are simply ${\rm d}H=0$, describing pure gauge configurations, while the boundary equations 
reproduce \eqref{int bd eq},
where $f(Y)={\partial g(Y)}/{\partial Y}$ is an anti-selfdual $(p+1)$-form function of a selfdual variable $Y=H^+$.

The action \eqref{CSbInt} describes arbitrary Abelian interacting theories of a single chiral $2k-$form field in $d=4k+2$ dimensional spacetime (the boundary $\partial M$) endowed with a metric of Lorentzian signature. 

In six dimensions, there is a unique functionally independent scalar made of a selfdual 3-form, therefore, \eqref{CSbInt} describes an infinite number of consistent theories parameterized by a function of one variable \cite{Avetisyan:2022zza}. In ten and higher dimensions such theories are parametrized by a function of more than one variable, as many as the number of independent Lorentz scalars constructed from a selfdual form. In two dimensions, there is no polynomial scalar constructed from a selfdual vector, therefore the only option of the form \eqref{CSbInt} is the free Abelian theory. For multiple fields, however, interactions via bulk non-Abelian deformations are possible \cite{Arvanitakis:2022bnr}.

\section{Democratic description for $p$-forms}

We will use now the same logic to derive democratic Lagrangians for arbitrary $p$-forms (including arbitrary Abelian interactions from \cite{Avetisyan:2022zza}). The starting point is the topological theory given by the action (occasionally referred to as the BF theory)
\begin{align}
    S_{\rm\sst Bulk}=\int_M (-1)^{d-p}\,G\wedge {\rm d}F+{\rm d}G\wedge F\,,
    \label{bulk act}
\end{align}
where $M$ is a $(d+1)$-dimensional manifold with $d$-dimensional boundary, $F$ is a $(p+1)-$form and $G$ is a $(d-p-1)-$form. 
Here, both $d$ and $p$ are arbitrary, as opposed 
to the previous section.
The gauge symmetry is given by
\begin{align}
    \delta F={\rm d}\alpha\,,\quad \delta G={\rm d}\beta\,.
\end{align}
The Lagrangian is gauge invariant up to boundary terms.
The bulk equations of motion are ${\rm d}F=0={\rm d}G$, implying that these fields are pure gauge, therefore there are no bulk degrees of freedom. The boundary term in the variation of the bulk Lagrangian is given by $\int_{\partial M}\delta G\wedge F- G\wedge \delta F\,.$
Adding to the action
\eqref{bulk act} a boundary term,
\begin{align}
-\int_{\partial M} \frac12(F\wedge\star F+G
\wedge\star G)\,,
\label{bd term}
\end{align}
modifies the boundary variation as
\begin{align}
    \int_{\partial M}\delta F\wedge ((-1)^{p+d+ pd}\,G-\star F)
    +\delta G\wedge (F-\star G)\nonumber\\
=(-1)^{p+d+pd}\int_{\partial M}\star \delta(F+\star G)\wedge (F-\star G)\,.
\end{align}
Here, again, we take the Neumann boundary condition $F-\star G=0$,
which can be viewed as 
the dynamical equations with respect
to the boundary variation,
so that the variational principle
gives the equations ${\rm d}F=0={\rm d}G$ supplemented with these boundary conditions.
The boundary term \eqref{bd term} again uses
a metric with Lorentzian signature.

Generalization to the self-interacting case is given as
\begin{align}
    S=\int_M (-1)^{d-p}\,G\wedge {\rm d}F+\dd G\wedge F\qquad\qquad\nonumber\\
-\int_{\partial M} \frac12\,(F\wedge\star F+G\wedge \star G)+g(F+\star G)\,,\label{intdemoc}
\end{align}
which gives the 
same bulk equations $\dd F=0=\dd G$ and the following modified boundary conditions:
\begin{align}
    F-\star G= f(F+\star G)\,.
\end{align}
Here again, $f(Y)={\partial g(Y)}/{\partial Y}$ for a $(p+1)-$form argument $Y$. This reproduces the democratic theory of general Abelian self-interactions for $p$-forms (the reduction to the democratic Lagrangians of \cite{Avetisyan:2022zza} will be demonstrated below).

An interesting observation \cite{Pulmann:2019vrw} is that, as opposed to the chiral case, now we also have the option to describe the boundary theory in a non-democratic manner by simply integrating out one of the fields. E.g., we can solve the bulk equation for $G$, that is $\dd F=0$, which implies $F=\dd A$. Substituting this into the action reduces the whole system to a boundary Lagrangian that is algebraic in $F=\dd A$, while the only field variable is now $A$. In the case of free theory, we will simply get a Maxwell Lagrangian $F \wedge \star F$.
Instead, for nontrivial $g(Y)$, we get a nonlinear algebraic equation expressing $G$ in terms of $F$, similar to those discussed in \cite{Avetisyan:2022zza,Avetisyan:2021heg}. Such relations are not always easy to solve explicitly even for nonlinear electrodynamics in $3+1$ dimensions, where some simplifications occur compared to general $d$ and $p$. These equations, however, explicitly capture the essence of the conversion procedure between democratic and ordinary single-field formalisms.
Note that we could equally integrate out $F$ instead of $G$ arriving at different but equivalent $d$-dimensional descriptions.
The two theories, corresponding to
two different reductions (either integrating out $G$ or $F$), are related by duality \cite{Pulmann:2019vrw}.
This is somewhat 
similar to the dualization
procedure where
we integrate out the field $A$ and $F$ from the action $S=\int_{\partial M}
-\frac12\,F\wedge \star F+
G\wedge (F-\dd A)$.
In the non-Abelian case,
this procedure leads to non-polynomial action in terms of the variable $G$,
with no smooth free limit \cite{Fradkin:1984ai}.

The democratic action
\eqref{intdemoc} for $p=2k$-forms in $d=4k+2$ dimensions can be diagonalized by introducing new variables $C=(F+G)/\sqrt{2}$ and $D=(F-G)/\sqrt{2}$
as
\begin{align}
    S=\int_M C\wedge \dd C-D\wedge \dd D\qquad\qquad\qquad\qquad\nonumber\\
-\int_{\partial M} \frac12\,(C\wedge\star C+D\wedge\star D)+g(C_++D_-)\,,
\label{CD act}
\end{align}
thus explicitly describing one chiral and one antichiral $p$-forms. 
Note that
the Abelian interaction
term $g(C_++D_-)$
can be viewed as a function
of two independent variables $C_+$ and $D_-$,
which are simply
the selfdual and anti-selfdual projections of $C_++D_-$, which means that \eqref{CD act} actually represents the most general interactions for one chiral and one antichiral fields $C$ and $D$.

Note that the normalization of the fields in the democratic setup is not unique: one can rescale the fields $F$ and $G$ in an opposite manner, arriving at the action,
\begin{align}
    S=&\ \int_M (-1)^{d-p}\,G\wedge {\rm d}F+\dd G\wedge F\qquad\qquad\nonumber\\
    &-\int_{\partial M} \bigg[ \,\frac12\,(\lambda^{-2}\,F\wedge \star F+\lambda^2\,G\wedge \star G)\nonumber\\
   &\hspace{40pt} +g(\lambda^{-1}\,F+\lambda\,\star G)\,\bigg]\,,
\end{align}
with boundary equations of motion,
\begin{align}
    \dd F=0=\dd G\,,\quad \lambda^{-1}\,F-\lambda\,\star G=f(\lambda^{-1}\,F+\lambda\,\star G)\,.
\end{align}
When coupled to charged matter (see for example \cite{Lechner:1999ga}), this rescaling is related to the change in the coupling constant, which requires opposite rescaling for electric and magnetic couplings.
This rescaling freedom is consistent with the Dirac-Schwinger quantization of the charges since the product of their coupling constants is invariant (the quantization applies only to the linear combination of pairwise product of electric and magnetic charges).


\subsection{Nonlinear electrodynamics and $SO(2)$ duality}

When $d=4k$, and both $F$ and $G$ are $p+1=2k$-forms, it is convenient
to label them as $F=H^1$ and $G=H^2$\,.
The Abelian nonlinear $p$-form theory in the democratic form, given in \cite{Avetisyan:2021heg}, can be derived from a $d+1=4k+1$-dimensional topological action with a boundary term,
\begin{align}
    S=&\int_M \epsilon_{bc}\, H^b\wedge \dd H^c \nonumber \\
    &- \int_{\partial M} \frac12\,H^b \wedge \star H^b+g(\star H^b+\epsilon^{bc}H^c)\,.
    \label{HH act}
\end{align}
This action transmutes under the reduction procedure of \cite{Arvanitakis:2022bnr} to that of \cite{Avetisyan:2021heg}.

The function $g(Y)$
is further restricted \cite{Avetisyan:2021heg} if we require
the $SO(2)$ duality symmetry rotating $H^1$ and $H^2$.
When $d=4$, the duality-symmetric theories of nonlinear electrodynamics are given by the five-dimensional action of type \eqref{HH act}
where the Abelian interaction 
term is reduced to a function 
of a single variable, $g(W^{ab}\,W_{ab})$.
Here, $W^{ab}$ is the duality covariant Lorentz scalar,
$$W^{ab}=\star[(\star H^a+\epsilon^{ac}H^c)\wedge\star (\star H^b+\epsilon^{bd} H^d)]\,,$$
whose trace vanishes identically: $W^{a}{}_a=0$\,.
The next example is  $d=8$, where the interactions in the general democratic 3-form theory will be parameterized by a function of $14$ variables, two for each order in fields --- from second to eighth. The duality-symmetric condition leaves only half of these variables --- seven: one for each order.

\section{Reduction to boundary theories}

We now proceed to the dimensional reduction procedure introduced in \cite{Arvanitakis:2022bnr} to show that the action \eqref{CSbInt} can be reduced to the nonlinear chiral $p$-form actions of \cite{Avetisyan:2022zza}. For that, one introduces a closed one-form $v$ (and corresponding vector which we will denote with the same letter) and decomposes the bulk field as:
\begin{align}
    H=\hat H+v\wedge \check H\,,
\end{align}
with a gauge redundancy
\begin{equation}
    \delta \hat H=-v\wedge \alpha\,,\qquad \delta \check H=\alpha\,,
\end{equation}
which was fixed by the choice $i_v \hat H=0$ in \cite{Arvanitakis:2022bnr}.
Plugging this decomposition into the Lagrangian, we notice that the field $\check H$ becomes a Lagrange multiplier enforcing a constraint on the field $\hat H$,
\begin{align}
    v\wedge \dd \hat H=0\,,
\end{align}
which can be solved following the Appendix C of \cite{Bansal:2021bis}, arriving at
\begin{align}
    H = \dd A+v\wedge R\,,
\end{align}
where $A$ and $R$ are $p$-forms. 
Then, one can see that the bulk Chern-Simons term of the action becomes a total derivative taking into account that $\dd v=0$. Therefore, the full action reduces to a bulk terms contribution to the boundary $\dd A\wedge v \wedge R$ plus boundary term, where the field $H$ is replaced by $\dd A+v \wedge R$.
Thus the final boundary action is given as
\begin{align}
    S=\int_{\partial M} -\frac12\,H\wedge \star H+\dd A\wedge v\wedge R+g(\star H + H)\,,\label{nonlinearchiralform}
\end{align}
where $H=\dd A+v\wedge R$. 

The equation \eqref{nonlinearchiralform} reproduces the Lagrangian for the arbitrary interacting theory of chiral $p$-form given in \cite{Avetisyan:2022zza} with one small difference: there, the $v$ is parameterized as $v=\dd a$ with a dynamical field $a$, thus avoiding the need for a prescribed one-form in the theory that naively breaks the Lorentz symmetry. The shift symmetry of the field $a$, which we call henceforth `PST symmetry' due to its close relation to the similar symmetry featured in the PST theory \cite{Pasti:1996vs}, is hard to anticipate from the Chern-Simons point of view.\footnote{Naively, in order to get the boundary Lagrangian, one needs to use a specific $v$. However, any non-null $v$ gives a consistent theory on the boundary, and all such theories are equivalently encoded in the action \eqref{CSbInt} which has manifest Lorentz symmetry. This gives an intuitive picture of why there should be extra gauge symmetries in the boundary theory that provide for Lorentz invariance, as in \cite{Pasti:1996vs,Pasti:1997gx,Buratti:2019guq,Avetisyan:2022zza,Evnin:2022kqn,Avetisyan:2021heg,Bansal:2021bis}, though it is not obvious how to make these symmetries explicit in the bulk theory language.} This symmetry, however, is crucial for the consistency of the theory and furthermore makes it possible to gauge-fix the field $a$ to a non-dynamical fixed function, at the expense of manifest Lorentz symmetry (thus making contact with the Chern-Simons derivation above). One may add a top-form term $J\wedge \dd v$ to the Lagrangian (where $J$ is a Lagrange multiplier)
and keep the field $v$ unconstrained.
This formulation (for the free theory) was the starting point in \cite{Mkrtchyan:2019opf} (where the one-form $v$ was denoted as $c$). 
Note, that the condition $v^2\neq 0$ is essential for the theory given by action \eqref{nonlinearchiralform} to describe a chiral form. One way to exclude the space $v^2=0$ from the theory could be an extra condition $v^2=1$ imposed by a Lagrange multiplier $\mu$, i.e., adding\footnote{We thank Chris Hull for discussions on this matter.} a term  $\mu (v^2-1)$ to the Lagrangian \eqref{nonlinearchiralform}.

Within the boundary theory, the expression $\star H+H$ is gauge-invariant with respect to the enlarged set of gauge symmetries shifting the auxiliary fields \cite{Avetisyan:2022zza}. Thus, these gauge symmetries guide us to the action \eqref{nonlinearchiralform} in the language of the boundary theory of  \cite{Avetisyan:2022zza}, while in the Chern-Simons language, the structure of the corresponding boundary terms is guessed so that they give rise to self-interacting chiral edge modes.

Now that we reviewed the derivation of \cite{Arvanitakis:2022bnr} and generalized it to include Abelian interactions of chiral forms, we will proceed to the democratic formulation for arbitrary $p$-forms. Using 
the same reduction procedure as in the chiral case, one can show that \eqref{intdemoc} leads to the general Abelian self-interactions for the $p$-forms, with the democratic boundary Lagrangian given in \cite{Avetisyan:2022zza}.
For that, one decomposes the fields $F$ and $G$ using a closed one-form $v$ (and corresponding vector which we will denote with the same letter):
\begin{equation}
    F=\hat F+v \wedge \check F
    \,,\qquad 
    G=\hat G+v\wedge \check G
    \,.
\end{equation}
Substituting this in the bulk Lagrangian, we can see that the fields $\check F$ and $\check G$ are Lagrange multipliers, imposing the constraints on the fields $\hat F$ and $\hat G$,
\begin{align}
    v\wedge \dd \hat F=0=v\wedge \dd \hat G\,,
\end{align}
which can be solved as earlier.

Substitution of the latter expressions in the action leads to purely boundary theory with a Lagrangian,
\begin{align}
    {\cal L}\,&=v\wedge S\wedge \dd A- \dd B\wedge v\wedge R \nonumber \\
    &\quad +\frac12\,(F\wedge \star F+G\wedge \star G)+g(\star G+F)\,,
\end{align}
where $H_1$ and $H_2$ are given by 
\begin{align}
    F=\dd A+v\wedge R\,,\label{H1}\\
    G=\dd B+v\wedge S\,.\label{H2}
\end{align}
This Lagrangian
coincides with \cite{Avetisyan:2022zza} after solving the constraint $\dd v=0$ as $v=\dd a$ and a simple field redefinition discussed in \cite{Bansal:2021bis}.

\section{Bulk-induced interactions}

The interactions introduced above only enter the higher-dimensional topological description through the boundary terms. Consequently, the interactions
in the resulting boundary theory are expressed through the field strength alone, but not through the gauge potential. It is possible to construct
more general interactions by considering topological interactions in the bulk. The simplest example of such interactions 
would be the non-Abelian Chern-Simons Lagrangian discussed in \cite{Arvanitakis:2022bnr}. More generally, one can add bulk interaction terms that are top-form wedge products of the fields involved. Such interactions are very limited for a single field, which we will discuss here, completing the discussion on Abelian self-interactions, and leaving the less constrained cases with multiple fields for future work.

For the chiral case, the only field is the $(p+1)-$form $H$, so the interactions may have the form $H\wedge H\wedge H$. Such a term is only legitimate in three bulk dimensions, where $H$ is a one-form, and even there, it is trivial for a single field $H$. For higher dimensions, self-interactions of a single chiral field can only be introduced via the boundary terms discussed earlier.

For democratic fields, the situation is different. In special cases, there is a possibility to add interacting terms for a single field. This happens when $d=3p+2$ for odd $p$, and the corresponding bulk term is $F\wedge F\wedge F$ (we recall that $F$ is a $(p+1)-$form and therefore the latter term is nontrivial for odd $p$ and is a top form in $d+1=3(p+1)$ dimensions). Therefore, the full action is given as
\begin{align}
S=&\ \int_M G\wedge {\rm d}F+\dd G\wedge F+\frac23\,\lambda_3\, F\wedge F\wedge F\nonumber\\
&-\int_{\partial M} \frac12\,(F\wedge\star F+G\wedge \star G)+g(F+\star G)\,.\label{intcs}
\end{align}
In the first non-trivial case, $p=1$, the $\lambda_3$ term in the action \eqref{intcs} describes Abelian Chern-Simons interactions for five-dimensional nonlinear electrodynamics. This can be quickly verified by integrating out the field $G$, most easily done in the case $g(Y)=0$, leading to Maxwell-Chern-Simons theory.

In the next case, $p=3$, the $\lambda_3$ term describes the Chern-Simons interactions for the three-form in eleven dimensions. This interaction is essential for the 11d supergravity and was the missing element for the democratic formulation of the latter in the same line as type II supergravities in ten dimensions \cite{Mkrtchyan:2022xrm}.

More generally, bulk Abelian interactions are possible in the dimensions $d=np+n-1$ (assuming that $p$ is odd) and are given by a wedge product of $n$ copies of $F$. For the quartic interactions, the first nontrivial case is the seven-dimensional Abelian Chern-Simons term, given by the bulk interaction $\lambda_4\, F\wedge F\wedge F\wedge F$.

The reduction procedure of \cite{Arvanitakis:2022bnr} works smoothly also in the presence of the bulk interaction \eqref{intcs}. The same procedure as performed above in the case of $\lambda_3=0$ leads to a neat cancellation of all bulk terms and leaves a boundary theory with the Lagrangian,
\begin{align}
{\cal L}=\ &v\wedge S\wedge \dd A- \dd B\wedge v\wedge R-\frac{\lambda_3}{3}A\wedge \dd A\wedge \dd A \nonumber \\
& +\frac12\,(F\wedge \star F+G\wedge \star G)
    +g(\star G+F)\,,\label{AB}
\end{align}
where $F$ takes the same form as in  \eqref{H1}
while $G$ is modified to
\begin{equation}
    G=\dd B+v\wedge S-\lambda_3\, A\wedge \dd A\,.
\end{equation}
This Lagrangian describes democratically nonlinear Maxwell-Chern-Simons theory in five dimensions for 1-form $A$ and 2-form $B$. The same Lagrangian describes democratically the 3-form $A$ in eleven-dimensions on equal footing with its dual 6-form $B$. 

\section{Maximal Supergravities in $d=10, 11$}

We can now quickly derive the type II supergravities in the democratic form of \cite{Mkrtchyan:2022xrm} from a topological theory in eleven dimensions. The starting point is the Chern-Simons action on the 11-dimensional manifold $M$ with a Lorentzian $10d$ boundary $\partial M$,
\begin{align}
    S_{\rm\sst RR}=\int_M G\wedge DG +\int_{\partial M} \frac12 (G, \star G)\,,\label{11dCS}
\end{align}
where $\star$ is defined with a factor $\star\alpha=(-1)^{\left\lfloor \frac{\deg\alpha}{2}\right\rfloor+\deg\alpha}*\alpha$ compared to Hodge star denoted in this section as $*$, and we use Mukai pairing $(\alpha,\beta):=(-1)^{\left\lfloor \frac{\deg\alpha}{2}\right\rfloor}(\alpha\wedge\beta)^{\on{top}}$, and finally $D=\dd+H\wedge$, where $H$ is a closed 3-form curvature of the Kalb-Ramond field (see details in \cite{Mkrtchyan:2022xrm}).

Here, $G$ encodes all the curvatures of RR fields:
\begin{align}
    G&=G_2+G_4+G_6+G_8+G_{10},\qquad \text{(IIA case)}\\
    G&=G_1+G_3+G_5+G_7+G_9.\phantom{\;\;}\qquad \text{(IIB case)}
\end{align}
The action \eqref{11dCS} can be reduced to ten dimensions via the procedure of \cite{Arvanitakis:2022bnr} to reproduce the RR sector actions of \cite{Mkrtchyan:2022xrm}. It is straightforward to add the NSNS sector and gravity, which are not described democratically.

An analogous description can be proposed for the 11-dimensional supergravity \cite{Cremmer:1978km}. Here, we introduce a 12-dimensional BF theory with a 11-dimensional boundary term and describe democratically the 3-form field with 4-form curvature $F$ and its dual 7-form curvature $G$ of the 6-form potential. Therefore, the action
takes the form of \eqref{intcs}
where the coupling constant is
fixed by supersymmetry as $\lambda_3=1$,
whose value is responsible for the remarkable exceptional symmetries of the dimensional reductions of $11d$ supergravity \cite{Henneaux:2015opa}. 
When $g(Y)=0$, we can integrate out the $G$ field from \eqref{intcs} to recover 
the standard 11d action
 involving a single three-form potential field.
Instead, if we reduce the $12d$ action \eqref{intcs} via the procedure of \cite{Arvanitakis:2022bnr},
 we find the democratic description 
 of the $11d$ Lagrangian
 of the form \eqref{AB} (with $\lambda_3=1$).

Integrating out the auxiliary fields $R$ and $S$, we recover the PST form of the action from \cite{Bandos:1997gd}. Note that deformations similar to $\alpha'-$corrections in String Theory are suggested by a non-trivial interaction term
$g(\star G+F)$.

\section{Discussion}

We have provided a simple derivation of arbitrary self-interacting Abelian $p$-form theories with first-order equations of motion --- democratic or chiral --- starting from familiar topological theories, making use of the ideas introduced in \cite{Arvanitakis:2022bnr}.
We also introduced large classes of Abelian self-interactions for these fields. The last missing piece of the puzzle was the Abelian interactions that cannot be written in terms of curvatures and are given by Abelian Chern-Simons terms that are only gauge invariant up to boundary terms. This setup builds a connection between Lagrangian formulations for the nonlinear (twisted) selfduality equations \cite{Avetisyan:2022zza} and other influential considerations in the literature (see, e.g. \cite{Floreanini:1987as,Henneaux:1988gg,Tseytlin:1990nb,Schwarz:1993vs,McClain:1990sx,Devecchi:1996cp,Bengtsson:1996fm,Rocek:1997hi,Bekaert:1999sq,Roiban:2012gi} for a sample of historical references).
More general interactions between multiple different fields will be studied systematically elsewhere.

The topological description of 
the RR fields in ten-dimensional supergravities discussed in this letter also
provides supporting explanations 
on the resolution \cite{Mkrtchyan:2022xrm,Kurlyand:2022vzv}
of the puzzles of supergravity on-shell actions 
\cite{Kurlyand:2022vzv},
which have to be contrasted with the expectations from holography. 
This resolution, which does not rely on a specific vacuum solution, is made 
at the level of the democratic $d$-dimensional Lagrangians with a unique $(d-1)$-dimensional boundary term protected by the PST symmetry.
From the perspective of the $(d+1)$-dimensional topological theories, this boundary term 
lives on the boundary of the boundary,
and hence it is not
surprising that any ambiguity in such a term
is resolved. We expect that the analogous puzzle of $11d$ supergravity
related to the electric solution \cite{Beccaria:2023hhi}
admits a similar resolution.

The democratic descriptions discussed here require a Lorentzian metric on the boundary
because the (twisted) self-duality equations 
with signature $(t,d-t)$
admit non-trivial solutions only for $+1 (-1)$ values of the Hodge star squared
$\star^2=(-1)^{p(d-p)+t}$.
Gravitational theories involving such actions may use path integral over the metric with arbitrary signature (see for example \cite{Witten:2021nzp}). Then, the degrees of freedom described by the democratic (or chiral) formulations of $p$-forms will be switched off in even-time signatures, going to a lower-dimensional phase space compared to the Lorentzian signature.

\vspace{.2cm}

\section*{Acknowledgements}
\noindent We are grateful to Alex Arvanitakis, Chris Hull, Massimo Porrati, Arkady Tseytlin, and Fridrich Valach for helpful discussions, and Zhirayr Avetisyan, Calvin Chen, Lewis Cole, and Alexander Sevrin for feedback on the manuscript. 
O.\,E.\ is supported by Thailand
NSRF via PMU-B (grant numbers B01F650006 and
B05F650021). 
E.\,J.\ was supported by the National Research Foundation of Korea (NRF) grant funded by the Korea government (MSIT) (No. 2022R1F1A1074977).
K.\,M.\ was supported by the
European Union’s Horizon 2020 Research and Innovation
Programme under the Marie Skłodowska-Curie Grant
No.\ 844265, UKRI and STFC Consolidated Grant ST/T000791/1.

\appendix

\bibliography{main}

\end{document}